\documentclass[fleqn,usenatbib]{mnras}

\usepackage{newtxtext,newtxmath}

\usepackage[T1]{fontenc}

\DeclareRobustCommand{\VAN}[3]{#2}
\let\VANthebibliography\thebibliography
\def\thebibliography{\DeclareRobustCommand{\VAN}[3]{##3}\VANthebibliography}

\usepackage{graphicx}	
\usepackage{amsmath}	

\title[Magnetic-field-induced inspiral of binaries]{Magnetic-Field-Induced Inspiral of Binaries with Circumbinary Disk: Black Hole and Protostellar Systems}

\author[T. Matsumoto et al.]{
Tomoaki Matsumoto,$^{1}$\thanks{E-mail: matsu@hosei.ac.jp}
Kenta Hotokezaka,$^{2,3}$
and 
Kohei Inayoshi$^{4}$
\\
$^{1}$Faculty of Sustainability Studies, Hosei University, Fujimi, Chiyoda-ku, Tokyo 102-8160, Japan\\
$^{2}$Research Center for the Early Universe, Graduate School of Science, The University of Tokyo, Bunkyo, Tokyo 113-0033, Japan\\
$^{3}$Kavli IPMU (WPI), UTIAS, The University of Tokyo, Kashiwa, Chiba 277-8583, Japan\\
$^{4}$Kavli Institute for Astronomy and Astrophysics, Peking University, Beijing 100871, China
}

\date{Accepted XXX. Received YYY; in original form ZZZ}

\pubyear{2026}

\begin{document}
\label{firstpage}
\pagerange{\pageref{firstpage}--\pageref{lastpage}}
\maketitle

\begin{abstract}
The orbital decay of binary systems is a critical process for understanding the evolution of massive binary black holes (MBBHs) and binary star formation. Performing high-resolution three-dimensional magnetohydrodynamic (MHD) simulations, we investigate a binary system that accretes gas from an infalling envelope analogous to the collapse of molecular cloud cores in the context of binary star formation. Our simulations reveal the presence of outflows/jets launched from both the circumstellar (mini) disks and the circumbinary disk (CBD). The magneto-rotational instability is also excited within the CBD. These magnetic processes efficiently transport angular momentum in the gas surrounding the binary and thereby drive orbital decay, while a purely hydrodynamical model exhibits orbital expansion. The decay rate reaches $\sim 0.3-0.7\%$ per orbital period, depending on the  initial magnetic field strength. By appropriately scaling these numerical results, we propose a new mechanism for MBBHs mergers within a Hubble time, overcoming the bottlenecks encountered at separations near the final parsec scales. Additionally, we present a formation scenario for close twin binary star systems, emphasizing the significant role of magnetic processes in driving orbital evolution across various astrophysical systems.
\end{abstract}




\section{Introduction} \label{sec:intro}

The orbital evolution of binary systems plays a crucial role in the formation of binary stars and massive binary black holes (MBBHs).
It has been suggested equal-mass binaries (twin binaries) exhibit an excess in the mass-ratio distribution not only for low-mass stars but also for massive stars \citep{Moe17}. A recent analysis of Gaia data further reveals an overabundance of twin binaries across a wide range of separations (\citealt{El-badry19}, see also \citealt{Lucy06,Krumholz07,Raghavan10}).  This excess implies that a binary formation mode in which binary stars accrete gas from a common circumbinary disk (CBD) can drive the mass ratio toward unity, as numerous simulations have shown \citep{Bate97,Young15,Young15b,Matsumoto19}.

Observed CBDs typically have a scale of $\sim 100$~au \citep{Hioki07,Fukagawa13,Dutrey14,Takakuwa17,Alves19,Takakuwa20}. Thus, the disk-binary interaction may affect the formation of binary systems with separations from $\sim 100$~au down to $\sim 0.1 - 1$~au. However, the process of orbital decay of binaries to form close twin binaries is still under debate \citep{Offner23}. 

The disk-binary interaction is also of great interest in the evolution of MBBHs in galactic nuclei. 
Several Pulsar Timing Array (PTA) projects, NANOGrav \citep{Agazie23a}, EPTA and InPTA \citep{Antoniadis23}, PPTA \citep{Reardon23}, and CPTA \citep{Xu23}, have reported evidence for a gravitational-wave background (GWB). Its spectral shape is consistent with a GWB produced by inspirals of MBBHs with masses of $10^8$--$10^9M_{\odot}$. 
However, it remains unclear how MBBHs evolve to the orbits in the PTA bands, orbital periods of years, 
known as the final parsec problem \citep{Begelman1980}. 
The disk-binary interaction may play a role in the orbital decay of MBBHs.

The interaction between a circumbinary disk and a binary system has been extensively studied \citep[see the review by][]{Lai22}, and numerous numerical simulations have shown that the binary orbit tends to expand \citep[e.g.,][]{Satsuka17, Moody19, Munoz19}. \citet{Tiede20} and \citet{Heath20} found that orbital decay can occur when the gas temperature is sufficiently low, and \citet{Franchini21} demonstrated that gas self-gravity can also reduce the binary separation. \citet{DOrazio21}, \citet{Dittmann22}, and \citet{Siwek23} conducted wide-range parameter surveys of orbital evolution, varying the binary eccentricity, gas temperature, and viscosity. These studies were carried out within the framework of hydrodynamics.

One of the key processes promoting orbital decay is angular momentum transport, for which magnetic fields serve as a powerful agent. Several studies have considered magnetic fields in binary accretion models \citep[e.g.,][]{Noble12, Shi12, Shi15, Bowen18, Lopez-Armengol21, Noble21, Avara24, Matsumoto24, Most24}. \citet{Shi12} estimated a slow decay rate of $\dot{a}_b/a_b = -8\times10^{-4} (M_d/M_b) \Omega_b$, where $a_b$, $M_d$, $M_b$, and $\Omega_b$ denote the binary separation, disk mass, binary mass, and binary orbital frequency, respectively. More recently, \citet{Matsumoto24} demonstrated that outflows driven by magnetic fields and the magneto-rotational instability (MRI) efficiently remove angular momentum from binary systems. \citet{Most24} reported orbital decay resulting from the long-term evolution of binary stars over $\sim200$ orbital periods.

Most of these studies assumed that the initial magnetic field configuration was confined within the circumbinary disk, such that the closed magnetic field lines followed the isodensity surfaces of the disk. However, in the context of binary star formation under typical magnetic field strengths, a natal molecular cloud core collapses along the field lines, and the resulting disk is threaded by open magnetic fields \citep[e.g.,][]{Matsumoto17, Tsukamoto18}. In addition, accretion onto the binary system from the cloud core plays an important role \citep{Bate97,Bate00}.

In this paper, we propose a new scenario for rapid binary orbital decay driven by magnetic fields and accretion. Our model incorporates the orbital evolution of binary stars and demonstrates that the rate of orbital decay depends on the magnetic field strength.
In contrast to most previous studies, we consider (1) an initial configuration in which the magnetic field is open, and (2) gas accretion onto the binary system from an infalling envelope. The latter process is analogous to the model of \citet{Bate97}, corresponding to accretion from a natal molecular cloud core in the context of binary star formation, and to high accretion triggered by a major merger of gas-rich galaxies in the case of massive binary black holes (MBBHs).

This paper is organized as follows. In Section~\ref{sec:models_methods}, we present the model and methods. The results of the simulations are presented in Section~\ref{sec:results}. Finally, the summary and discussion of this paper are given in Section~\ref{sec:summary_discussion}.

\begin{table*}
  \caption{Model parameter $v_A$ and resulting orbital parameters $\dot{a}_b$ and $e_b$}
  \label{table:model-parameters}
\begin{tabular}{lrrrr}
\hline
Name &
$v_A/v_0$ &
$\left[(\dot{a}_b /a_b)T_b\right]_\mathrm{ave}$ &
$e_{b, \mathrm{ave}}$ &
$\left[\ell /(G M_b r_b)^{1/2} \right]_\mathrm{ave}$ \\
\hline
Hydrodynamical model & 0.0  & $0.00122$  & 0.0398 & 0.449 \\
Fiducial field model & 0.1  & $-0.00771$ & 0.0445 & 0.147 \\
Strong field model   & 0.5  & $-0.00361$ & 0.0441 & 0.189 \\
\hline
\end{tabular}
\\
\footnotesize{
The third column represents the average values of $(\dot{a}_b /a_b)T_b$ over the time interval $t/T_0 \in [12, 20]$.
The fourth column represents the eccentricity of the binary orbit averaged over $t/T_0 \in [1, 20]$.
The fifth column represents the eivenvalue $\ell$ averaged over $t/T_0 \in [15, 20]$.
}
\end{table*}

\begin{figure*}
\includegraphics[width=\textwidth]{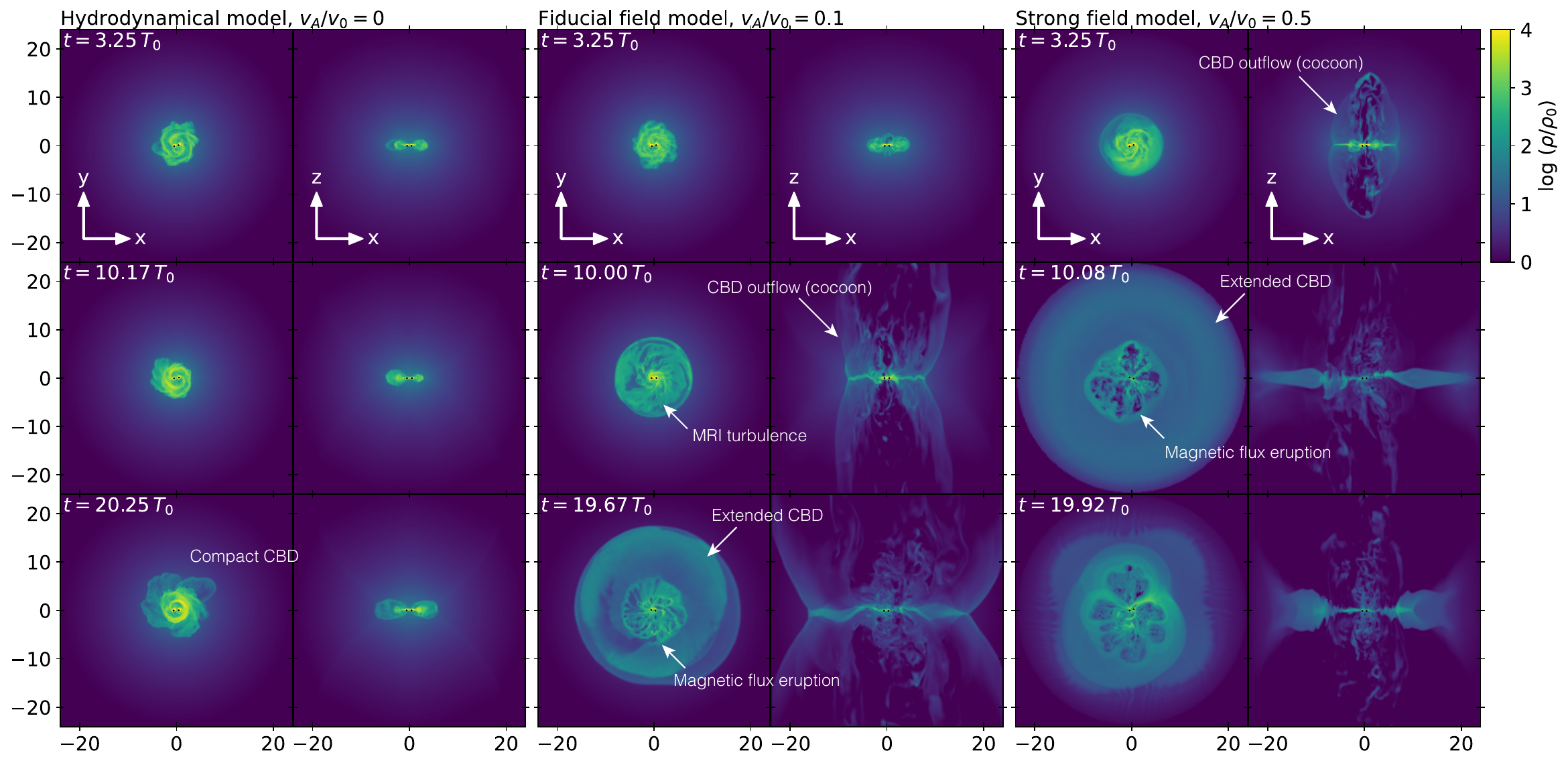}\\
\includegraphics[width=\textwidth]{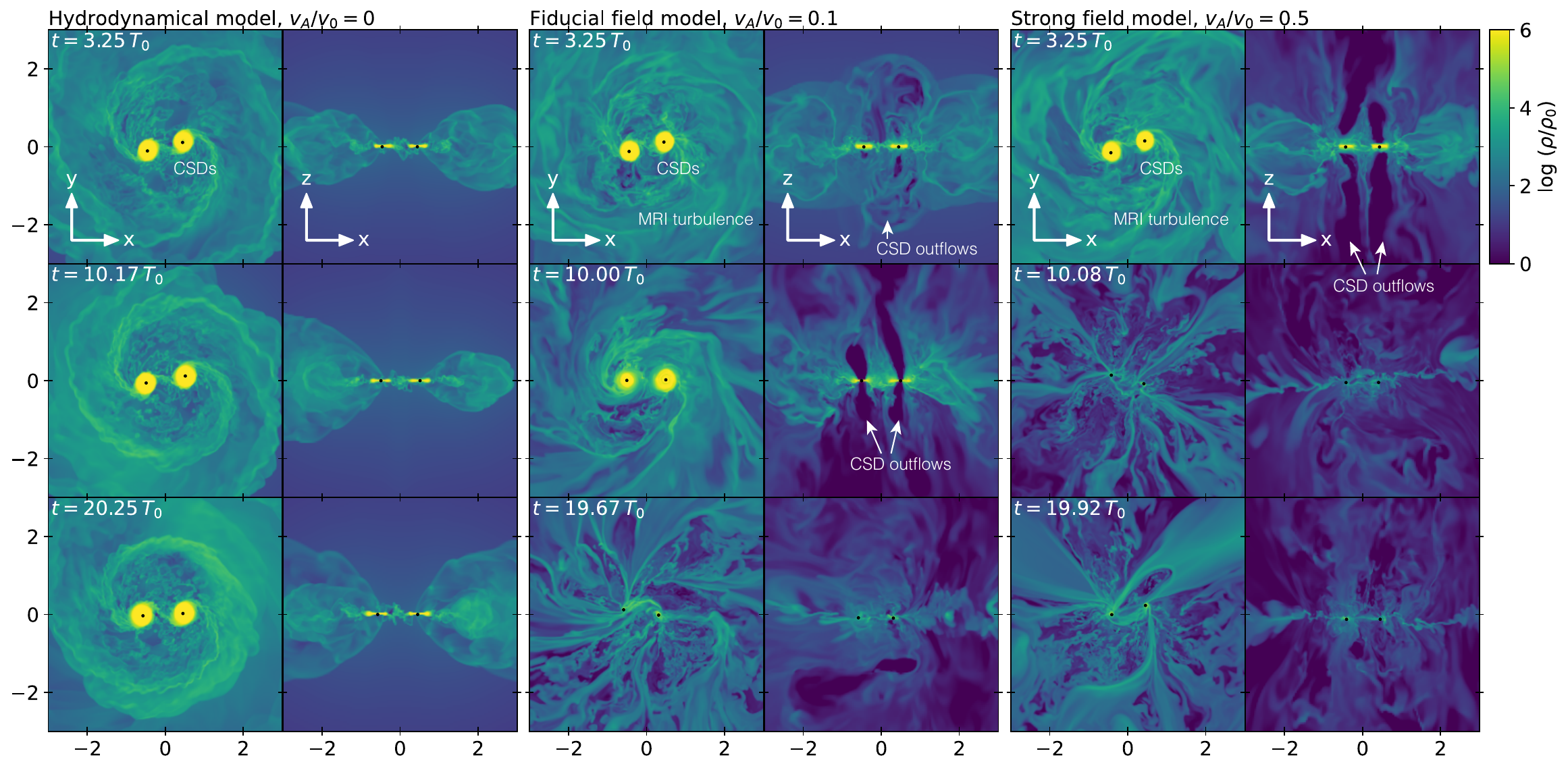}
\caption{Evolution of the hydrodynamical model ($v_A/v_0 = 0$; left column), 
the fiducial field model ($v_A/v_0 = 0.1$; middle column), 
and the strong field model ($v_A/v_0 = 0.5$; right column), 
at $t \simeq 3 T_0$ (upper row), $10 T_0$ (middle row), $20 T_0$ (bottom row). 
For each model, the left and right panels show the logarithmic density distributions in the $z=0$ and $y=0$ planes, respectively. The black dots represent the sink particles.
Upper panels show a wide region in the plotted plane ($48 a_0 \times 48 a_0$), while lower panels show the central region close to the binary stars ($9 a_0 \times 9 a_0$).
\label{timeseries_models_lev1.pdf}}
\end{figure*}

\section{Models and Methods}
\label{sec:models_methods}
The present models extend our previous work by \citet{Matsumoto24} to include the orbital evolution of binary stars. We consider a binary system orbiting around the origin of the computational domain, which has a cylindrical shape with a radius of $L$ ($= 24 a_0$) and a height of $[-192 a_0, 192 a_0]$, where $a_0$ is the initial binary separation. Gas is injected through the cylindrical boundary surfaces because we consider here a binary system accreting gas from an infalling envelope. The gas density at the boundary is prescribed as $\rho(r) = \rho_0 (r/L)^{-3/2}$, where $r$ denotes the spherical radius.
The injected gas has a radial velocity of $v_r = [2 G M_0 / r - (\Omega_\mathrm{inf} R)^2]^{1/2}$ at the boundary, corresponding to free-fall from infinity. Here, $R$, $M_0$ ($= M_{1,0} + M_{2,0}$), $M_{1,0}$, $M_{2,0}$, and $\Omega_\mathrm{inf}$ denote the cylindrical radius, the initial total mass of the binary, the initial masses of the primary and secondary stars, and the angular velocity of the infalling gas, respectively.
We assume a rigidly rotating, infalling envelope characterized by a given radial scale $r$. The angular velocity is taken to be a function of $r$, expressed as $\Omega_\mathrm{inf} = j_\mathrm{inf}/r^2$, where $j_\mathrm{inf}$ is the specific angular momentum of the gas accreting onto the midplane, set to $j_\mathrm{inf} = 1.2(G M_0 a_0)^{1/2}$.
The density normalization at the boundary is $\rho_0 = 7.62\times10^{-7} M_0 a_0^{-3}$, corresponding to a gas injection rate of $\dot{M}_\mathrm{env} \sim 0.01 M_0 / T_0$, where $T_0 = 2\pi (a_0^3 / G M_0)^{1/2}$ is the initial orbital period of the binary. An isothermal equation of state with a sound speed $c_s$ is adopted.

In the case of binary star formation, the gas injection rate assumed above corresponds to
\begin{equation}
\dot{M}_\mathrm{env} \sim 10^{-5}\,M_\odot \mathrm{yr}^{-1} \left(\frac{M_0}{M_\odot}\right)^{3/2} \left(\frac{a_0}{100\,\mathrm{au}}\right)^{-3/2},
\end{equation}
which is consistent with the typical accretion rate of protostars \citep{Shu87,Dunham14}. In the case of MBBHs, the gas injection rate in units of the Eddington limit is expressed as
\begin{equation}
\frac{\dot{M}_\mathrm{env}}{\dot{M}_\mathrm{Edd}} \sim \frac{\tau_\mathrm{Sal}}{100T_0} = 85.4\left(\frac{M_0}{10^{8.5}M_\odot}\right)^{1/2}\left(\frac{a_0}{1\,\mathrm{pc}}\right)^{-3/2}, 
\end{equation}
where $\dot{M}_\mathrm{Edd}$ is the Eddington accretion rate, and $\tau_\mathrm{Sal} = M_0/\dot{M}_\mathrm{Edd}\simeq 45\,\mathrm{Myr}$ is the Salpeter timescale assuming a radiative efficiency of 10\%.
The mass injection rate adopted here is larger than the accretion rates typically inferred for AGNs, which are usually a fraction of the Eddington accretion rate \citep{Heckman14,Inayoshi20}. The relatively large injection rate is adopted to clearly measure the orbital evolution within a limited simulation time. Since both the BH mass accretion rate and the gravitational interaction between the BHs and the CBD scale approximately with the amount of gas supplied to the system, the orbital evolution rate also roughly scales with the gas injection rate. Therefore, the results presented here can be rescaled to systems with lower accretion rates.

The simulation accounts for the gravitational interaction between the binary stars and the gas, as well as the mutual gravity between the two stars, while the self-gravity of the gas is neglected. Gas structures such as spiral arms formed in the CBD can therefore exert gravitational torques on the binary. The motion of the stars (modeled as sink particles) is evolved under these gravitational forces together with the mass and momentum accreted from the gas. 

The initial condition is constructed based on the results of a hydrodynamical simulation with a fixed circular orbit and fixed masses of binary stars in $-10 T_0 \le t \le 0$. Then, a uniform magnetic field $B_{z,0}$ is imposed, and an ideal MHD simulation starts at $t = 0$. For $t > 0$, the orbital evolution and mass growths of the binary stars are accounted for. 

The calculations were performed using \texttt{SFUMATO} \citep{Matsumoto07}, utilizing fixed mesh refinement with grid levels of $\ell = 0$ to 6. The corresponding numerical resolutions are $\Delta x_\mathrm{max} = 0.375 a_0$ (for $\ell = 0$) to $\Delta x_\mathrm{min} = 0.00586 a_0$ (for $\ell = 6$). An isothermal version of the Boris-HLLD scheme \citep{Matsumoto19HLLD} is adopted. The solver has third-order accuracy in space using the MUSCL method and second-order accuracy in time with the predictor-corrector method. Sink particles are used as a model of the binary stars. The sink particles accrete the mass and momentum of the gas exceeding a threshold density $\rho_\mathrm{sink} = 10^6 \rho_0$, within the sink region with a radius $r_\mathrm{sink} = 4 \Delta x_\mathrm{min} = 0.0234 a_0$. The accretion affects the orbital evolution of the sink particles as described above. The detailed prescription of the sink particles is presented in \citet{Matsumoto15}.

We assume ideal MHD throughout the computational domain, except for the sink particle implementation where a strong Ohmic dissipation is introduced inside the sink region in order to decouple the gas from the magnetic field within the sink radius. This treatment prevents the rotating gas inside the sink region from artificially launching outflows. As discussed in \citet{Matsumoto24}, the outflows appearing in the simulation are instead driven by the circumstellar disks (CSDs; or mini disks), not by numerical effects associated with the sink particles.

In this paper, we focus on the cases of twin binary models ($q=M_{2,0}/M_{1,0} = 1$) and isothermal sound speed of $c_s = 0.1 v_0$, where $v_0$ ($=(GM_0/a_0)^{1/2}$) is the initial orbital velocity of binary stars. Three models are examined, changing the initial strength of magnetic field $B_{z,0}$, and corresponding Alfv\'en speed of envelope gas $v_A$ ($= B_{z,0} /\left(4\pi \rho_0 \right)^{1/2}$) as shown in Table~\ref{table:model-parameters}. 
The calculations were terminated at $t \sim 20 T_0$, when the outflows reached the top and bottom boundaries at $z = \pm 192 a_0$. This ensures that the outflows did not reflect at the boundaries, which would otherwise affect the subsequent evolution of the binary system.

\section{Results}
\label{sec:results}

\subsection{Evolution of models}

Figure~\ref{timeseries_models_lev1.pdf} depicts the time evolution of the three models. For the models with magnetic fields, two outflows, which appear as low-density regions, are launched from the CSDs surrounding each star. In addition, an outflow exhibiting a cocoon-like shape is launched from the CBD. The magneto-rotational instability (MRI) occurs in the CBD, showing turbulent density structure in the CBD. This results in the redistribution of angular momentum and leading to the CBD expansion. As a result, these magnetic phenomena extract angular momentum from the system in both the radial and vertical directions. 

\begin{figure}
\includegraphics[width=\columnwidth]{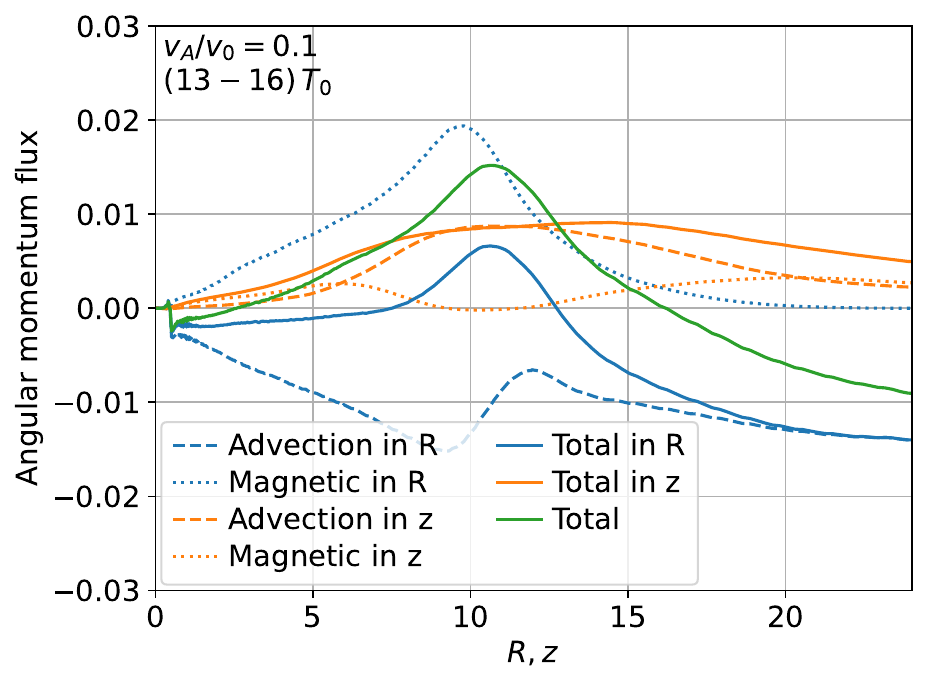}
\caption{
Angular momentum flux transported through a cylindrical surface of radius $R$ and height $\pm z$ for the fiducial model, averaged over  $t = (13-16)T_0$. The green line shows the total flux, while the blue and orange lines show the components in the $R$- and $z$-directions, respectively. Each component is further separated into contributions from gas advection (dashed) and magnetic fields (dotted). The plotted quantity represents the total angular momentum passing through the surface, rather than a flux per unit area. Positive values indicate outward transport, while negative values indicate inward transport.
\label{j_flux_binobmhd_q1j12c01va01h.4968700.pdf}
}
\end{figure}

Figure~\ref{j_flux_binobmhd_q1j12c01va01h.4968700.pdf} shows the angular momentum transfer for the fiducial model in a quantitative manner. The outflow transports angular momentum outward in the vertical direction; this vertical transport is partly carried by gas advection associated with the outflows (orange dashed line) and partly by magnetic braking (orange dotted line). In the radial direction, the magnetic field transports angular momentum outward mainly through MRI (blue dotted line), while the infalling envelope carries angular momentum inward (blue dashed line).
This behavior of angular momentum transport is consistent with the results reported in \citet{Matsumoto24}.

We also note that the CSDs become faint in the course of the simulations and disappear at the late stages (Figure~\ref{timeseries_models_lev1.pdf}). This fading of the CSDs is attributable to the decreased accretion onto the binary stars and their CSDs, as later quantified in Figure~\ref{plot_time_mdot_rbin_models.pdf}.
We also observe magnetic flux eruptions from sink particles, which manifest as holes in the CBDs. This phenomenon has also been reproduced in magnetically arrested disk (MAD) simulations \citep{Igumenshchev08, Tchekhovskoy11, Most24}.
Although the sink particle implementation adopted in those studies differs from that used here, the appearance of magnetic flux eruptions in different numerical setups suggests that this phenomenon may not depend sensitively on the specific sink particle treatment.

Magnetic flux eruptions modify the local gas dynamics around the sink particles, and therefore may affect the momentum carried by the gas that is subsequently accreted onto the sink particles. However, within the duration of the present simulations, we do not observe clear abrupt changes in the sink particle motion that are temporally correlated with the occurrence of magnetic flux eruptions. Previous studies indicate that magnetic flux eruptions occur intermittently \citep[e.g.,][]{Tchekhovskoy11}, and their possible long-term influence on binary orbital evolution remains an interesting subject for future investigation.

In contrast to the MHD models, due to the absence of MRI, the hydrodynamical model exhibits a more compact CBD than the magnetic models. The two spiral arms are more prominent compared to the magnetic models because the turbulent level is lower. An asymmetry with an $m=1$ mode is seen in the CBD, as reported by \citet{Matsumoto19}.

\subsection{Orbital evolution of binary stars}

\begin{figure}
\includegraphics[width=0.93\columnwidth]{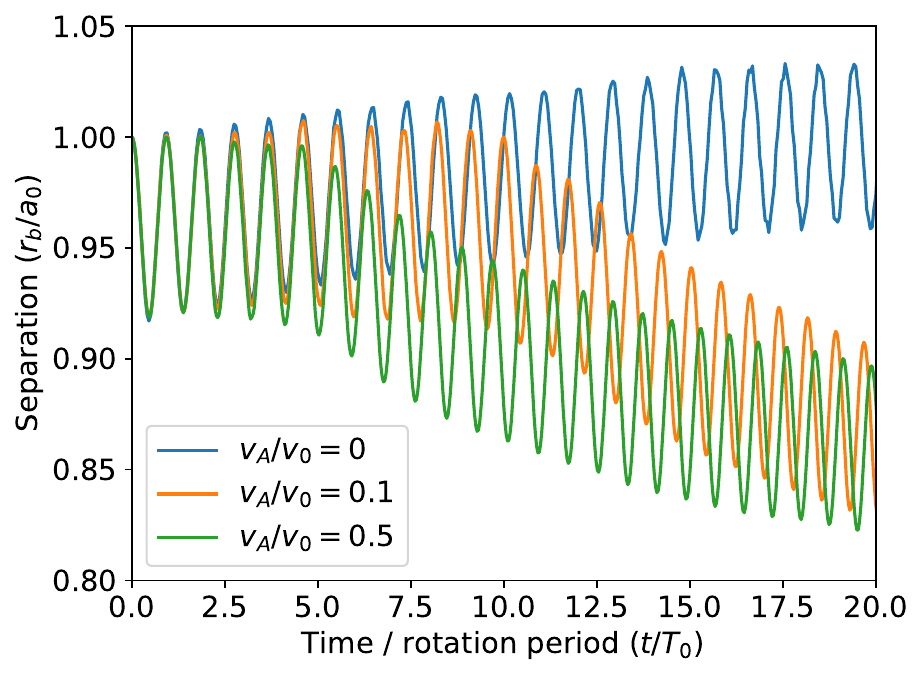}
\;\;\;\;\;\; \includegraphics[width=\columnwidth]{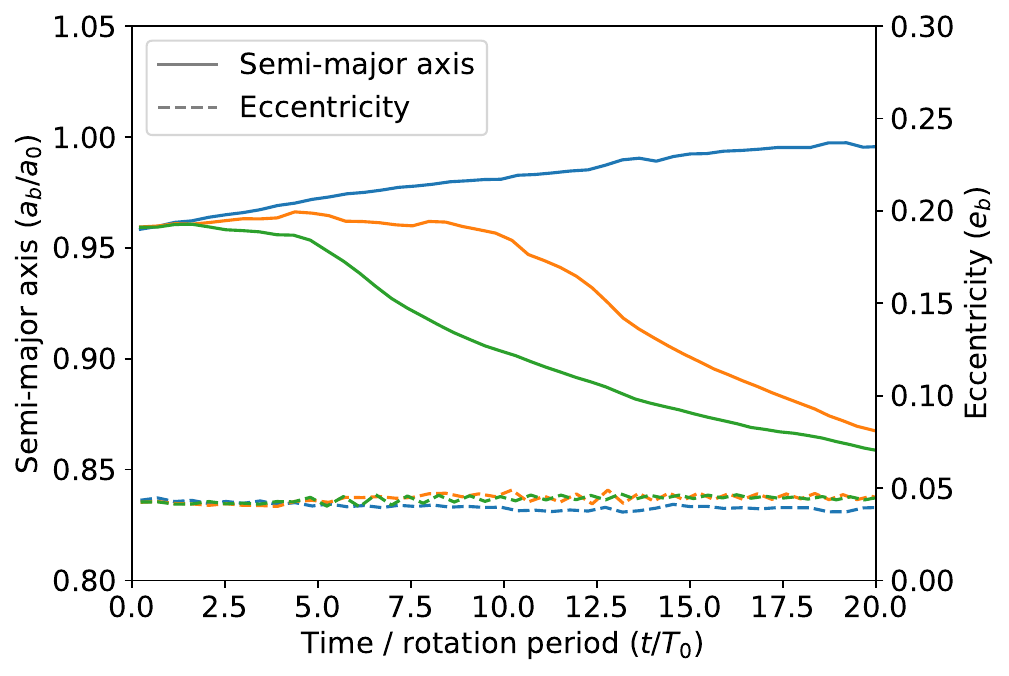}
\caption{
Binary separation (upper panel) and semi-major axis and eccentricity (lower panel) as a function of time. In the lower panel, solid lines (left axis) show the semi-major axis, and dashed lines (right axis) show the eccentricity.
 \label{plot_time_pr_models.pdf}}
\end{figure}

Figure~\ref{plot_time_pr_models.pdf} (upper panel) shows the separation of binary stars $r_b$ as a function of time. While the separation increases with time in the hydrodynamical model ($v_A = 0$),  it decreases in the two models with magnetic fields, indicating that the magnetic fields significantly affects the binary orbital evolution.
Figure~\ref{plot_time_pr_models.pdf} (lower panel) shows the semi-major axis $a_b$ and eccentricity $e_b$ calculated from the binary separation $r_b$. 
The semi-major axis and eccentricity are measured as $a_b = (r_{b,\mathrm{max}} + r_{b,\mathrm{min}})/2$ and $e_b = (r_{b,\mathrm{max}} - r_{b,\mathrm{min}})/(r_{b,\mathrm{max}} + r_{b,\mathrm{min}})$, respectively, where $r_{b,\mathrm{max}}$ and $r_{b,\mathrm{min}}$ are the local maxima and minima in the time evolution of $r_b$ (corresponding to apastron and periastron). These quantities represent instantaneous orbital elements defined from the radial extrema.
The semi-major axis begins to decrease earlier in the model with the strong magnetic field than in the fiducial field model. After that, the semi-major axis continues to decrease in both the fiducial and strong field models.

The change rate of the semi-major axis normalized by the orbital period $(\dot{a}_b/a_b) T_b$ is shown in Table~\ref{table:model-parameters}, where $T_b$ is the orbital period measured from the time evolution of $r_b$, and $\dot{a}_b$ is estimated from the time evolution of $a_b$ using finite differences. The change rate is measured in the late phase of the evolution, and it exhibits positive and negative values for hydrodynamical and MHD models, respectively. The strong field model has a lower negative value ($\sim -0.003$) than the fiducial model ($\sim -0.007$) because the decrease in the semi-major axis gradually slows in the strong field model, as shown in Figure~\ref{plot_time_pr_models.pdf} (lower panel).

The oscillation of the separations in Figure~\ref{plot_time_pr_models.pdf} (upper panel) is attributed to the eccentricity of the orbits. As shown in Figure~\ref{plot_time_pr_models.pdf} (lower panel)  and Table~\ref{table:model-parameters}, the eccentricity remains small in all the models, indicating that the binary stars maintain nearly a circular orbit.

\begin{figure}
\includegraphics[width=\columnwidth]{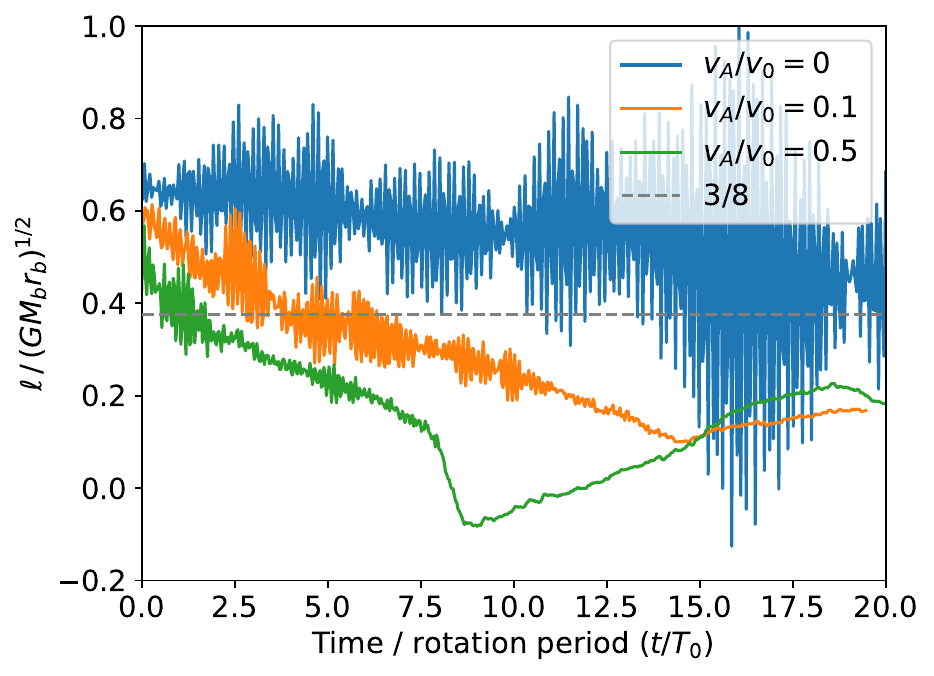}
\caption{The specific angular momentum eigenvalue $\ell = \dot{J}_b/\dot{M}_b$ as a function of time. 
The bifurcation point where the orbit expands or decays is $\ell = 3/8 (GM_br_b)^{1/2}$ (dashed line).
Moving averages with a window size of $4T_0$ are taken to observe long-term trends.
\label{plot_time_ell0_models_0.pdf}
}
\end{figure}

\subsection{Eigenvalue of specific angular momentum}

Figure~\ref{plot_time_ell0_models_0.pdf} shows the evolution of the ``eigenvalue'' of specific angular momentum $\ell = \dot{J}_b/\dot{M}_b$ as a function of time, where $J_b$ and $M_b $ denote the total angular momentum and mass of the binary stars, respectively\footnote{In this paper, the symbol $\ell$ is employed for the eigenvalue of the specific angular momentum, whereas $\ell_0$ is used for it conventionally. The subscript 0 is utilized for the initial condition in this paper.}. To estimate the eigenvalue, we monitored the mass and linear momentum accretions and gravitational torques on the sink particles during the simulation.

The eigenvalue is an indicator of the evolution of the binary orbit \citep{Miranda17,Munoz19,Lai22}. When assuming an equal mass binary, $J_b = (M_b/4) a_b^2 \Omega_b$ and $\dot{J}_b = \dot{M}_b \ell$ lead to the following equation,
\begin{equation}
\frac{\dot{a}_b}{a_b} = 8\left(\frac{\ell}{a_b^2 \Omega_b} - \frac{3}{8}\right)\frac{\dot{M}_b}{M_b}.
\label{eq:eigenvalue}
\end{equation}
Since the binary orbits remain nearly circular ($e_b \lesssim 0.05$), we use Equation (\ref{eq:eigenvalue}) as a good approximation to describe the orbital evolution.
Equation~(\ref{eq:eigenvalue}) indicates that the binary orbit decays when $\ell < (3/8) a_b^2 \Omega_b \simeq  (3/8) (GM_b r_b)^{1/2}$.
As expected from equation~(\ref{eq:eigenvalue}), binary orbital decay occurs when the eigenvalue matches this criterion (when $t/T_0 \gtrsim 4$ for the fidicual model, and $t/T_0 \gtrsim 1$ for the strong field model) (see Figure~\ref{plot_time_ell0_models_0.pdf} and Figure~\ref{plot_time_pr_models.pdf} (bottom)).
Table~\ref{table:model-parameters} also shows the average values of the eigenvalues. For the magnetic models, the eigenvalues are $\ell \sim (0.15 - 0.19)(G M_b r_b)^{1/2}$ for the relatively late stages, and $\ell = (0.16 - 0.18)(G M_b r_b)^{1/2}$ for the end of the calculations.

In contrast, the hydrodynamical model exhibits an $\ell$ greater than $(3/8) (G M_b r_b)^{1/2}$, except for an oscillation with a high-amplitude regime. This is consistent with the continuous increase in the semi-major axis.

\begin{figure}
\includegraphics[width=\columnwidth]{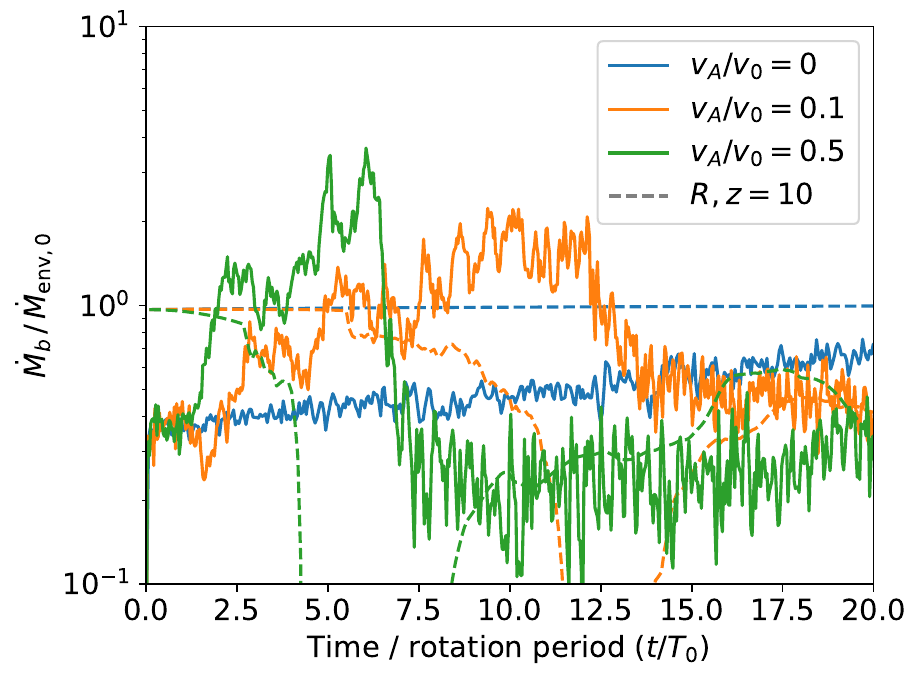}
\caption{
Accretion rates onto the binary stars (solid lines) as functions of time. For each model, the total accretion rate $\dot{M}_b$ ($= \dot{M}_1 + \dot{M}_2$) is shown, normalized by the initial accretion rate of the envelope. The dashed lines show the accretion rates measured on the cylindrical surfaces with radius $R = 10 a_0$ and hight $z= \pm 10 a_0$.
\label{plot_time_mdot_rbin_models.pdf}
}
\end{figure}

Equation~(\ref{eq:eigenvalue}) indicates that the evolution of the binary separation is regulated by the accretion onto the binary stars. Figure~\ref{plot_time_mdot_rbin_models.pdf} shows that the magnetic models exhibit a high accretion rate in the early phase, but it decreases to $\dot{M}_b/\dot{M}_\mathrm{env,0} \sim 0.2$–$0.6$ in the late phase. This reduction in the accretion rate results in a slight slowdown of the orbital decay, as shown in the lower panel of Figure~\ref{plot_time_pr_models.pdf}. 
The reduction in the accretion rate is caused by the outflow from the CBD. As shown in Figure~\ref{timeseries_models_lev1.pdf}, the outflow sweeps away the infalling envelope. This interaction lowers the accretion rate at $r, z = 10$, as indicated by the dotted lines in Figure~\ref{plot_time_mdot_rbin_models.pdf}, and consequently leads to a decrease in the accretion rate onto the binary stars.

\section{Summary and Discussion}
\label{sec:summary_discussion}

We conducted high-resolution 3D MHD simulations with fixed mesh refinement to investigate the impact of magnetic fields on the orbital decay of binaries, which rapidly accrete gas from an infalling envelope analogous to the collapse of molecular cloud cores in the context of binary star formation. 
In our numerical models with magnetic fields, outflows and MRI-driven turbulence extract angular momentum from gas at the central part and transport it outward. In contrast to the purely hydrodynamical case, where the binary semi-major axis expands, the inclusion of magnetic fields leads to binary orbital decay. 

Although the simulations do not reach a long-term steady state, the qualitative difference between the magnetized and non-magnetized models persists over multiple orbital periods. This suggests that magnetic effects play a robust role in the orbital evolution.

The typical decay rate of the semi-major axis follows Equation~(\ref{eq:eigenvalue}) with the eigenvalue of specific angular momentum $\ell$. 
When assuming $\ell = 0.17~a_b^2 \Omega_b$, consistent with our numerical result (see Figure~\ref{plot_time_ell0_models_0.pdf}), the orbital decay timescale is expressed as 
\begin{equation}
\tau_\mathrm{decay} = \frac{a_b}{|\dot{a}_b|} = \mathcal{A} \frac{M_b}{\dot{M}_b},
\label{eq:tau_decay_mdot}
\end{equation}
where $\mathcal{A}\sim 0.6$. 
Thus, the decay timescale is comparable to the mass accretion timescale. 
Under the outer boundary conditions in our simulations, where the mass injection rate from the envelope is $\dot{M}_\mathrm{env} \sim 0.01 M_0/T_0 $ (see Section~\ref{sec:models_methods}), and the mass accretion rate on to the binary stars is $\dot{M}_b \sim \mathcal{B} \dot{M}_\mathrm{env}$ where $\mathcal{B} \sim 0.4$ (see Figure~\ref{plot_time_mdot_rbin_models.pdf}), the decay timescale simplifies to
\begin{equation}
  \tau_\mathrm{decay} \sim 100~\frac{\mathcal{A}}{\mathcal{B}} T_0 \sim  \frac{2\pi}{0.007} \left(\frac{a_b^3}{GM_b}\right)^{1/2},
  \label{eq:tau_decay}
\end{equation}
indicating that the binary orbit shrinks at a rate of 0.7\% per orbital period. This estimate is consistent with our numerical results with magnetic models (see Table~\ref{table:model-parameters}).

For binary protostars with a separation of $a_0=50$ au and a total mass of $1\,M_{\sun}$, Equation~(\ref{eq:tau_decay}) yields an orbital decay timescale of $\tau_{\rm decay} = 5\times 10^4~\mathrm{yr}$, comparable to the typical protostar lifetime of $10^4$~yr. This mechanism may be responsible for the formation of close twin binaries \citep{El-badry19}.
For a wider separation of $a_0=100\,\mathrm{au}$ with the same total mass, Equation~(\ref{eq:tau_decay}) gives $\tau_\mathrm{decay} = 1.5\times 10^5\,\mathrm{yr}$, which is longer than the typical protostellar lifetime. This suggests that the efficiency of orbital decay during the protostellar phase is sensitive to the initial binary separation.

An intriguing application of our findings is the orbital evolution of MBBHs at parsec-scale separations.
This scale is crucial, as binary decay often stalls before reaching this point due to insufficient stellar and gaseous angular momentum transport, the so-called final parsec problem \citep[e.g.,][]{Begelman1980}. 
Our results suggest that in dense environments, binary decay proceeds at a rate comparable to the mass accretion rate. When BHs accrete at near the Eddington limit, $\dot{M}_{\rm Edd}$, the decay timescale adjusts to the Salpeter timescale, $\tau_{\rm Sal}\simeq 45~{\rm Myr}$.  Thus, the orbital decay time is 
\begin{align}
    \tau_{\rm decay} \approx 45\,{\rm Myr}\,\mathcal{A}f_{\rm duty}^{-1}\left(\frac{\dot{M_b}}{\dot{M}_{\rm Edd}}\right)^{-1},
\end{align}
where $f_{\rm duty}$ is the duty cycle of the accreting phase.
Since this is shorter or comparable to the typical AGN lifetime \cite[e.g.,][]{Martini04,Hopkins06}, sustained accretion could significantly accelerate binary evolution. 
For an accretion at the Eddington rate, the density of the infalling gas is set considerably lower than that assumed in the present model, i.e., $\rho_0 (\tau_\mathrm{acc}/\tau_\mathrm{Sal})= 4.8 \times 10^{-22}~\mathrm{g\,cm}^{-3}~(M_{\rm b}/10^{8.5}~M_{\sun})^{1/2} (a_b/1~\mathrm{pc})^{-3/2}$ at $L=24 a_0$, where $\tau_\mathrm{acc} = M_b/\dot{M}_b \sim (100/\mathcal{B}) T_0 = 1.3~{\rm Myr} \, (M_{\rm b}/10^{8.5}~M_{\sun})^{-1/2} (a_b/1~\mathrm{pc})^{3/2}$.
However, directly simulating MBBH orbital decay in 3D MHD over such long timescales remains computationally challenging.
We leave it for future investigation.


Dense gas environments are likely provided by major mergers of gas-rich galaxies, which trigger AGN ignition as observed in ultra-luminous infrared galaxies \citep[ULIRGs;][]{Veilleux02,Ichikawa14,Assef15,Tsai15}.
In such merging systems, massive BHs are delivered into the galactic nucleus (at least the sphere of influence radius, a few hundreds of parsec for a $10^{8.5}~M_\odot$ BH) through gravitational interaction with surrounding stars and gas across multiple scales \citep{Armitage02,Chapon13,Capelo15,Farris15}.
If the nuclear region remains both gas-rich and sufficiently magnetized, magnetic-field-induced orbital decay can drive MBBHs toward parsec-scale separations within ULIRG, leading to their eventual merger.
By equating the orbital-decay rates from magnetic-field-driven angular momentum transport and GW emission \citep{Peter64}, we find that 
the MBBH evolution is dominated by GW losses when
the gravitational-wave frequency, $f_{\rm GW}=2f_{\rm orb}$ ($= 2T_b^{-1}$), reaches 
\begin{equation}
    f_{\rm GW} \approx 2.6\times10^{-9}\,{\rm Hz}\,\mathcal{A}^{-3/8}f_{\rm duty}^{3/8}\left(\frac{\dot{M}_b}{\dot{M}_{\rm Edd}}\right)^{3/8}\left(\frac{\mathcal{M}}{10^{8.5}M_{\odot}}\right)^{-5/8},
\end{equation}
where $\mathcal{M}=(M_1M_2)^{3/5}/(M_1+M_2)^{1/5}$ is the chirp mass of a binary, $M_1$ and $M_2$ are the component masses.
For a MBBH with $\mathcal{M}\sim 10^{8.5}M_{\odot}$, this frequency is about the minimum frequency, above which the 15-yr PTA observations are sensitive.
Since ULIRGs are common and may serve as a primary channel for BH mass growth (the AGN activity peaks around $z\sim 2$; see \citealt{Delvecchio14}), they may  significantly contribute to the stochastic GWB \citep{Inayoshi18}.
We note that our result of the magnetic-field-driven orbital decay with  Eddington accretion suggests a flattened GWB spectrum compared to the canonical $f^{-2/3}$ power law for $f \lesssim 2.4\cdot 10^{-8}\,{\rm Hz}$, which can be tested with the future PTA data sets \citep[e.g.][]{Agazie23b}.

\section*{Acknowledgements}
Numerical computations were carried out on XC50 (ATERUI II) and XD2000 (ATERUI III) at Center for Computational Astrophysics, National Astronomical Observatory of Japan, and on Yukawa-21 at YITP in Kyoto University.

This research was supported in part 
by JSPS KAKENHI Grant Numbers
JP18H05437,
JP17K05394,
20H05639, 
20H00158,
JP23K03464,
23H01169, 23H04900, and
the JST FOREST Program (JPMJFR2136).
K.I. acknowledges support from the National Natural Science Foundation of China (12573015, W2532003, 1251101148, 12233001), 
the Beijing Natural Science Foundation (IS25003), and the China Manned Space Program (CMS-CSST-2025-A09).

\section*{Data Availability}

The data underlying this article will be shared on reasonable request to the corresponding author.



\bibliographystyle{mnras}
\bibliography{ms} 








\bsp	
\label{lastpage}
\end{document}